\documentclass[11pt]{article}                               

\usepackage{latexsym}
\usepackage{amssymb}

\def\AFOUR{%
\setlength{\textheight}{9.0in}%
\setlength{\textwidth}{5.75in}%
\setlength{\topmargin}{-0.375in}%
\hoffset=-.5in%
\renewcommand{\baselinestretch}{1.17}%
\setlength{\parskip}{6pt plus 2pt}%
}

\AFOUR                                           

\makeatletter
\def\section{\@startsection {section}{1}{\z@}{-3.5ex plus -1ex minus
 -.2ex}{2.3ex plus .2ex}{\large\bf}}
\def\subsection{\@startsection{subsection}{2}{\z@}{-3.25ex plus -1ex minus
 -.2ex}{1.5ex plus .2ex}{\normalsize\bf}}
\makeatother

\makeatletter
\@addtoreset{equation}{section}

\makeatother

\newcommand{\nc}{\newcommand}
\newcommand{\rnc}{\renewcommand}

\nc{\bea}{\begin{eqnarray}}
\nc{\eea}{\end{eqnarray}}
\nc{\be}{\bea}
\nc{\ee}{\eea}

\rnc{\a}{\alpha}
\nc{\ab}{\bar{\a}}
\nc{\ap}{\a^{+}}
\nc{\abm}{\ab^{-}}
\rnc{\b}{\beta}
\nc{\bb}{\bar{\b}}
\nc{\bbp}{\bb_{\zb}^{+}}
\nc{\bm}{\b_{z}^{-}}
\nc{\oa}{\overline{\a}}
\nc{\ob}{\overline{\b}}
\rnc{\gg}{\gamma}
\rnc{\d}{\delta}
\nc{\f}{\phi}
\nc{\fb}{\bar{\phi}}
\nc{\vf}{\varphi}
\nc{\p}{\psi}

\rnc{\c}{\chi}
\nc{\la}{\lambda}
\nc{\m}{\mu}
\nc{\n}{\nu}
\rnc{\o}{\omega}
\nc{\Om}{\Omega}
\rnc{\t}{\theta}
\nc{\eps}{\epsilon}
\rnc{\S}{\Sigma}
\nc{\F}{\Phi}
\nc{\ldb}{\left(\!\left(}
\nc{\rdb}{\right)\!\right)}

\nc{\trac}[2]{{\textstyle\frac{#1}{#2}}}

\nc{\ex}[1]{\mbox{e}^{\,\textstyle#1}}

\nc{\mat}[4]{\left(\begin{array}{cc}#1&#2\\#3&#4\end{array}\right)}

\nc{\som}[9]{\left(\begin{array}{ccc}#1&#2&#3\\#4&#5&#6\\#7&#8&#9%
\end{array}\right)}

\nc{\tr}{\mathop{\mbox{tr}}\nolimits}
\nc{\ad}{\mathop{\mbox{ad}}\nolimits}
\nc{\Tr}{\mathop{\mbox{Tr}}\nolimits}
\nc{\Det}{\mathop{\mbox{Det}}\nolimits}
\rnc{\det}{\mathop{\mbox{det}}\nolimits}
\nc{\rk}{\mathop{\mbox{rk}}\nolimits}
\nc{\ra}{\rightarrow}
\nc{\Ra}{\Rightarrow}
\nc{\LRa}{\Leftrightarrow}
\nc{\ot}{\otimes}
\rnc{\ss}{\subset}
\nc{\nul}{\noindent\underline}
\nc{\non}{\nonumber\\}

\nc{\subs}[1]{{\vspace*{0.5cm}}%
{\noindent\underline{#1}}{\addcontentsline{toc}{subsection}{#1}}%
{\vspace*{0.3cm}}}

\nc{\zb}{\bar{z}}
\rnc{\lg}{\mathfrak{g}}
\nc{\lt}{\mathfrak{t}}
\nc{\lk}{\mathfrak{k}}
\nc{\lh}{\mathfrak{h}}
\nc{\pik}{\Pi_{\lk}}
\nc{\pip}{\Pi_{+}}
\nc{\pim}{\Pi_{-}}
\nc{\pih}{\Pi_{\lh}}
\nc{\jz}{J_{z}}
\nc{\jzh}{\jz^{\lh}}
\nc{\jzp}{\jz^{+}}
\nc{\jzm}{\jz^{-}}
\nc{\del}{\partial}
\nc{\dz}{\del_{z}}
\nc{\dzb}{\del_{\bar{z}}}
\nc{\az}{A_{z}}
\nc{\azb}{A_{\bar{z}}}
\nc{\g}{g^{-1}}
\nc{\dw}{\Delta_{W}}
\nc{\Ad}{{\mbox{Ad}}}
\nc{\ks}{Ka\-za\-ma-\-Su\-zu\-ki}
\nc{\KS}{\ks}
\nc{\ksm}{\ks\ model}
\rnc{\AA}{{\Bbb A}}
\nc{\BB}{{\Bbb B}}
\nc{\CC}{{\Bbb C}}
\nc{\PP}{{\Bbb P}}
\nc{\cpm}{\CC\PP(m)}
\nc{\cpn}{\CC\PP(n)}
\nc{\cp}[1]{\CC\PP(#1)}
\nc{\gmn}{G(m,m+n)}
\nc{\gmnk}{\gmn_{k}}
\nc{\cO}{{\cal O}}
\nc{\bcO}{\bar{\cO}}
\nc{\bO}{\bar{O}}
\nc{\oQ}{\overline{Q}}

\begin{document}
\global\parskip=4pt

\makeatletter
\begin{titlepage}
\begin{center}
{\LARGE\bf 
Chern-Simons Theory on a General\\[.1in] Seifert 3-Manifold}\\
\vskip 0.7in
{\bf Matthias Blau}
\vskip .1in
Albert Einstein Center for Fundamental Physics\\ 
Institute for Theoretical Physics\\
University of Bern, Switzerland.
\vskip 0.2in
\vskip 0.2in
{\bf Keita Kaniba Mady}
\vskip .1in
University of Science-Techniques and Technology of Bamako, Mali \\
and Abdus Salam ICTP, 
 Trieste, 
Italy.\\
\vskip 0.2in
{\bf K.S. Narain \&  George Thompson}
\vskip .1in
Abdus Salam ICTP, 
 Trieste, 
Italy.\\

\end{center}
\vskip .4in
\begin{abstract}
\noindent
The path integral for the partition function of Chern-Simons gauge
theory with a compact gauge group is evaluated on a general Seifert
3-manifold. This extends previous results and relies on
abelianisation, a background field method and local application of the
Kawasaki Index theorem. 
\end{abstract}
\end{titlepage}
\makeatother

\begin{small}
\end{small}

\setcounter{footnote}{0}

\section{Introduction}

The main determination of the Reshetikhin-Turaev-Witten (RWT)
\cite{RT1, RT2, Witten}
invariants of a 3-manifold has been through the use of the
Reshetikin-Turaev construction or conformal field theory methods. A
sampling of these approaches is 
\cite{LR, Rozansky, Jeffrey, Hansen, Hansen-Takata}. There are also path
integral evaluations such as semi-classical evaluations \cite{Jeffrey}
as well as evaluations based on localisation \cite{BW, 
  Beasley} and those based on supersymmetric
localisation \cite{Khallen}. Though it must be said that the
localisation approaches 
are not exact (so far) in case there is a moduli space of flat
connections that is not made up of isolated points.

In a series of papers
\cite{BT-CS, btdia, BT-S1, BT-Seifert, BT-complex-CS} two of us introduced
the concept of diagonalisation as a gauge fixing condition in gauge
theories. If one starts with a trivial $G$-bundle over a manifold M\footnote{We 
have indicated how non trivial gauge 
  bundles can be incorporated in \cite{BHT}.}, with Lie algebra  
$\lg$, then in principle diagonalisation leaves one with a $\lt$ (some Cartan sub algebra)
bundle and associated vector bundles. That procedure requires, however, that the
3-manifold be a principal bundle or fibration (over an orbifold) and that one
make non-smooth gauge 
transformations to achieve the required gauge. The rationale for the first
requirement is that, as explained in
\cite{btdia}, this diagonalisation works ``best'' on 2-dimensional 
manifolds, since the resulting diagonalised gauge fields have precisely
the singularity structure that allows them to be interpreted as
non-singular connections on a non-trivial bundle. Generically in more
than 2 dimensions the required gauge transformations and resulting gauge
fields are too singular to lend themselves to such an intepretation,
and thus diagonalisation can only be applied if it is possible to
reduce the calculations to 2 dimensions.  In the case of 3-manifolds,
this is possible in principle if the 3-manifold has the structure of a
fibration over a 2-dimensional orbifold, to which the calculation can
be ``pushed down'', and this singles out Seifert 3-manifolds among all
possible 3-manifolds as those to which diagonalisation (at least as
understood by us at present) can be applied. 

A notable feature of this
approach to the calculation of the Chern-Simons partition function of 
Seifert 3-manifolds \cite{BT-Seifert} is that it completely bypasses
the (possibly arduous) task of having to integrate over some moduli
space of non-Abelian flat connections, as it essentially 
reduces the partition function to that of an Abelian gauge theory
on a 2-dimensional orbifold.

The singular
gauge transformations ``Abelianise'' the theory so that the fields are
well defined but are now sections of non-trivial Abelian bundles. The obstructions
\cite{btdia}
to using smooth gauge transformations to accomplish this are then 
reflected in the fact that one must
sum over the Abelian bundles that are generated in this way. In all of the cases considered
thus far the non trivial bundles that arise are always some power of a
fixed line bundle $\mathcal{L}_{M}$, over the orbifold base, depending
on the underlying
3-manifold $M$. Hence, there has only ever been the need to sum over one
integer (the first Chern class of $\mathcal{L}_{M}$) in the path
integral. The general class of Seifert three manifolds for which this
is true we dubbed $\mathbb{Q}$HS$[g]$ (genus $g$ generalisations of 
rational homology spheres) in \cite{BT-Seifert}. 

Our aim here is to extend the diagonalisation method to general
Seifert 3-Manifolds. In order to diagonalise on Seifert 3-manifolds,
which are not 
$\mathbb{Q}$HS$[g]$, requires some new techniques. Firstly, we note
that on a Riemann surface with $N$ orbifold points $\Sigma_{V}$ a general line
V-bundle may be decomposed as
\be
\mathcal{L} =\mathcal{L}_{0}^{n_{0}} \otimes \mathcal{L}_{1}^{n_{1}}
\otimes \dots \otimes \mathcal{L}_{N}^{n_{N}} 
\ee
with $0\leq n_{i} < a_{i}$ where $a_{i}$ is the order of the $i$'th
orbifold point while $\mathcal{L}_{0}$ is a smooth line bundle and
$n_{0} \in \mathbb{Z}$. By Theorem 2.3 in \cite{FS} for $M$ a smooth
Seifert 3-manifold
\be
\mathrm{H}^{2}\left(M, \, \mathbb{Z} \right) \simeq
\mathrm{Pic}^{\mathrm{t}}\left(\Sigma_{V}
\right)/\mathbb{Z}\left[\mathcal{L}_{M} \right] \oplus \mathbb{Z}^{2g}
\label{Pic}
\ee
where $\mathrm{Pic}^{\mathrm{t}}\left(\Sigma_{V}
\right)$ is the topological Picard group of topological isomorphism
classes of line V bundles over $\Sigma_{V}$. There is a more detailed
statement namely Proposition 5.3 in \cite{MOY} which explains the
relationship between bundles on $M$ and those on $\Sigma_{V}$. 

As all such bundles arise on diagonalisation
we will need, for each line bundle, in the gauge bundle, to sum over
the set of integers $(n_{0}, \, n_{i})$ (a $\rk{(G)}$'s set of such
integers for structure group $G$). Consequently we will need to
incorporate into the path integral that we are integrating
over connections on such 
non-trivial bundles. To do this we introduce a background connection
in anticipation that the connection is, infact, non-trivial. 

To describe the background connections in detail we need to explain
the orbifold construction on a Riemann surface and line V bundles in
some detail, the principal bundle structure of Seifert 3-manifolds and
the relationship between these. This is
done in Section \ref{Orbifold}, Section \ref{Seifert} and in Section
\ref{Holomorphic} respectively. One consequence of having an explicit
background connection is that one does not need to introduce such a
background implicitly in the evaluation of the determinants in Section
\ref{Phase} which is unlike the situation in the original evaluation
of such determinants given in \cite{BT-CS}. The reason for being so
explicit is that one needs to keep to the fore the fact that on
diagonalisation the smooth line bundles that are generated on the
3-manifold come from line V bundles below as essentially all the
calculations are done on the orbifold.

The calculational part rests in section \ref{calc}. The
original evaluation of the determinants in \cite{BT-CS} shows that
one is really dealing with densities on the underlying (V-)
surface. The background 
fields localise the calculations to their support. Once one realises
that the only changes that need to be made are to 
express the Kawasaki index theorem in a manner which takes into
account local information then the calculations in this paper become
essentially a commentary on \cite{BT-Seifert} explaining where
modifications need to be made, especially as we have alreay
incorporated the background connection. The only point to be aware of
is that we change our orientation and normalisation conventions in
section \ref{calc} to 
make it easier to use the results of \cite{BT-Seifert}.

\section{2-Dimensional Orbifolds and Seifert 3-Manifolds}\label{Orbifold}

For us a compact closed 2-dimensional orbifold or V manifold
$\Sigma_{V}$ is a genus $g$ Riemann surface with $N$ discs $D_{i}$
removed and replaced with the cones $U_{i} \simeq
D_{i}/\mathbb{Z}_{a_{i}}$ for $i=1, \dots, N$. The apex of
the cone is the orbifold point and we denote those points by
$x_{i}$. The local model is, for $z \in D_{i}$,
\be
z \simeq \zeta.z, \;\;\;\; \zeta \in \mathbb{Z}_{a_{i}}
\ee
so that local holomorphic coordinates on the $U_{i}$ are $z^{a_{i}}$
and we think of $\zeta$ as 
a complex $a_{i}$'th root of unity.

Complex line V bundles $\mathcal{L} \rightarrow \Sigma_{V}$ are
described in a similar fashion. Around an orbifold point the local
description is
\be
(z, w) \simeq (\zeta.z, \zeta^{b}.w), \;\; w \in \mathbb{C} \label{local-L}
\ee
where $0 < b < a$ and $\rho(\zeta) = \zeta^{b}$ is thought of as a
representation of $\mathbb{Z}_{a}$. We note that the circle V bundle
$S(\mathcal{L})$, with $|w|=1$ in (\ref{local-L}), is smooth as long
as the $\gcd{(a, b)}=1$ since there are no fixed points of the discrete
action $\zeta^{b}.w$ in this case (otherwise with $a= cd$ and $b= ce$
where $c>1$ one could take $\zeta = \exp{(2\pi i d/a)}$ so that $\zeta^{b}=1$).

Of special interest to us are the
building blocks of such bundles which we denote by $\mathcal{L}_{i}$.
The $\mathcal{L}_{i}$ are trivial outside of the local neighbourhood
$U_{i}$ and have 
local data on $D_{i} \times \mathbb{C}$
\be
(z, w) \simeq (\zeta_{i}.z, \zeta_{i}.w), \;\;\; \zeta_{i} \in
\mathbb{Z}_{a_{i}} 
\ee
Such holomorphic `point' V bundles can be described as follows \cite{FS}
\be
\mathcal{L}_{i} = \left( \Sigma_{i} \times \mathbb{C} \right)
\cup_{\psi} \left( D_{i} \times \mathbb{C}\right)/\mathbb{Z}_{a_{i}}
\ee
where $\Sigma_{i}$ is the smooth Riemann surface $\Sigma$ with $x_{i}$
removed and  the clutching map is defined, away from $z=0$, by
\be
\psi(z,w) = (z^{a_{i}}, z^{-1}.w)
\ee
and $\psi$ can be thought of as a $\mathbb{Z}_{a_{i}}$ invariant map
on $D/\{0\} \times \mathbb{C}$ which descends to $(D/\{0\} \times
\mathbb{C})/\mathbb{Z}_{a_{i}} $. The $n_{i}$-th tensor power of this
bundle, $\mathcal{L}_{i}^{\otimes n_{i}}$ has clutching map
\be
\psi(z,w) = (z^{a_{i}}, z^{-n_{i}}.w)\label{clutch}
\ee
A general holomorphic V bundle $\mathcal{L}$ over $\Sigma_{V}$ is then
obtained by performing this construction at each of the $N$ orbifold
points and at one regular point.

We are also interested in the unit disc V bundle $D(\mathcal{L})$ of
$\mathcal{L}$ which is obtained by taking $|w| \leq 1$ in
(\ref{local-L}) and which is realated to the circle V bundle
$S(\mathcal{L})$ by $\partial D(\mathcal{L}) =
S(\mathcal{L})$.

A Seifert 3-manifold $M[\deg{\left(\mathcal{L}_{M} \right)}, \, g, \,
(a_{1}, \, b_{1}), \dots , (a_{N}, \, b_{N}) ]$ is a smooth circle V
bundle $S\left(\mathcal{L}_{M}\right)$ over a genus $g$ Riemann surface with $N$
orbifold points with Seifert data $(a_{i}, \, b_{i})$ such that
\be
0 < b_{i} < a_{i}, \;\;\;\; \gcd{(a_{i}, \, b_{i})} =1
\ee
The first condition means that we are using normalised Seifert
invariants while the second is the condition that, as we saw, the
bundle is smooth.

Throughout we will have in mind a decomposition of the base space
$\Sigma_{V}$ into open sets $U_{i}$ for $i=0, 
1, \dots, N$ where $U_{0}= \Sigma_{0}$ and the
$U_{i}$ $i =1, \dots , N$ are the cones $D/\mathbb{Z}_{a_{i}}$ about
the orbifold points $x_{i}$, while $\Sigma_{0}$ is just $\Sigma_{V}$
with the cones excised and the line V-bundles $\mathcal{L}_{i}$ will
be `point' bundles localised on the $U_{i}$.

\subsection{Sections and Connections on V Bundles over $\Sigma_{V}$ }\label{Sections}

There is a natural section of the $\mathcal{L}_{i}^{\otimes n_{i}}$
namely on $D_{i}$ the section is
\be
s_{i}(z) = z^{n_{i}}
\ee
which can be extended over the rest of $\Sigma$ as the constant
section 1 via the clutching map (\ref{clutch}). The first Chern class
is
\be
c_{1}\left(\mathcal{L}_{i}^{\otimes n_{i}} \right) = \frac{n_{i}}{a_{i}}
\ee
A suitable local connection form on $D_{i}/\{ 0\}$ for
$\mathcal{L}_{i}^{\otimes n_{i}}$ is  
\be
\alpha_{i}^{\otimes n_{i}} = g(z\overline{z})\, d\ln{(z^{n_{i}})} +
\overline{w}dw \label{alphai} 
\ee
providing that $g$ is the identity much of the way into $D_{i}$ (we
always take the $D_{i}$ to be unit discs). Having such a $g$ is
consistent with the clutching map (\ref{clutch}). On $S\left(
  \mathcal{L}_{i}^{\otimes  n_{i}}\right)$ this is, with $w
= \exp{(i\sigma)}$, the connection 
\be
\alpha_{i}^{\otimes n_{i}} =  g(z\overline{z}) \, d\ln{(z^{n_{i}})} + id\sigma
\ee
so $d\alpha_{i}^{\otimes n_{i}}$ is horizontal and 
\be
d\alpha_{i}^{\otimes n_{i}} = n_{i}\, d\alpha_{i}
\ee
with holonomy
\be
c_{1}\left(  \mathcal{L}_{i}^{\otimes n_{i}}\right)=\frac{1}{2\pi i}
\int_{U_{i}} d\alpha_{i}^{\otimes n_{i}} = \frac{1}{2\pi i} 
\frac{1}{a_{i}} \int_{\partial D_{i}}
\alpha_{i}^{\otimes n_{i}} = \frac{n_{i}}{a_{i}} 
\ee
as required. The $\alpha_{i}$ are
also locally contact structures with
\be
\frac{1}{(2\pi i)^{2} }\int_{S\left( \mathcal{L}_{i}^{\otimes
      n_{i}}\right)} \alpha_{i} (\mathcal{L}_{i}^{\otimes n_{i}})
\wedge d\alpha_{i}(\mathcal{L}_{i}^{\otimes n_{i}}) = \frac{n_{i}}{a_{i}}
\ee

$\mathcal{L}_{i}$ is then the holomorphic line V bundle with first
Chern class $c_{1}(\mathcal{L}_{i})=1/a_{i}$ and with divisor at the
$i$'th orbifold point with $i \in 1,
\dots, \, N$ and allow, for $i=0$, $\mathcal{L}_{0}$ to be the line bundle at
a smooth point with first Chern class $c_{1}(\mathcal{L}_{0})=1$. Then
we have that any smooth holomorphic line V bundle $\mathcal{L}$ is given by
\be
\mathcal{L}= \mathcal{L}_{0}^{\otimes n_{0}} \otimes
\mathcal{L}_{1}^{\otimes n_{1}} \otimes \dots \otimes 
\mathcal{L}_{N }^{\otimes n_{N}}\label{line-bundle-decomp}
\ee 
with
\be
n_{0} \in \mathbb{Z}, \;\;\;\;  0< n_{i} < a_{i}, \;\;\;\; i =1, \dots , N 
\ee

\section{Surgery, Connections and Chern Classes}\label{Seifert}
This section is meant to connect the line bundle view point of the
previous section with the direct construction of the Seifert
3-manifold. We begin with a 
topological description of the circle V-bundles that we considered in the
previous section. This is followed by a surgery
prescription on glueing boundaries along tori relevant to creating
Seifert 3-manifolds.

\subsection{Solid Tori with $S^{1}$ Action of $(a,b)$ Type}
We fix an element of $SL(2, \mathbb{Z})$ in this section
\be
\left(\begin{array}{rr}
a & b \\
-r & -s
\end{array}
\right) , \;\;\; br =1 + as, \;\;\; \gcd{(a,b)}=1, \;\;\;  0<b<a,
\;\;\; 0 < r < a \label{solid}
\ee
Consider a solid torus $D^{2} \times S^{1}$, where $D^{2}$ is a unit
disc in $\mathbb{C}$ with center at the origin and with local coordinates
$(\rho e^{i\phi}, \, e^{i\psi})$. The standard $S^{1}$ action of type $(a,b)$
is
\be
\left(\rho\, \ex{i\phi}, \, \ex{i\psi}\right).\ex{i\theta} =
\left(\rho \, \ex{\left( i\phi +
    ir\theta\right)}, 
\, \ex{\left( i\psi + ia \theta\right)}\right) \label{S1-action}
\ee
We can quotient with this action (use $\theta$ to set $\psi=0$ and we
still have those transformations generated by $\zeta =
\exp{(i\theta)}$ where $\theta = 2\pi/a$ as 
these do not change the value of $\exp{(i\psi)}=1$) to be left with
$D^{2}/\mathbb{Z}_{a}$. Denote the solid torus with this action by
$V_{(a,b)}$ then we have the $S^{1}$ V-bundle $V_{(a,b)} \longrightarrow
D^{2}/\mathbb{Z}_{a}$.

The vector field corresponding to the generator of the $U(1)$ action
on $V_{(a,b)}$ is
\be
\xi = r\frac{\partial}{\partial \phi} +  a \frac{\partial}{\partial \psi} 
\ee
and the `vertical'dual one-form is
\be
d\theta= bd\phi - s d\psi \label{dual-1-form}
\ee
while the horizontal 1-forms, the space of which we quite generally
denote by $\Omega_{H}^{1}$, are spanned by
\be
d\rho, \;\;\; \mathrm{and} \;\;\; d\chi = ad\phi - r d\psi
\ee

\subsection{Surgery to obtain Seifert 3-Manifolds}
The exposition here partially follows that of Jankins and
Neumann \cite{JN} and of Orlik \cite{Orlik}.

A solid torus
is $D^{2} \times S^{1}$ where $D^{2}$ is a unit disc in $\mathbb{C}$
with center at the origin. Let $\lambda$ be
a longitude, that is a
simple non contractible curve on the $T^{2}$ boundary of $D^{2} \times
S^{1}$, and for
definiteness, fix the point $\{ 1\} \in \partial D$ and take $\lambda$
to be $\{ 1\}
\times S^{1}$. We also set $\mu$ to be a meridian, that is a contractible
loop in $D^{2} \times 
S^{1}$ lying on the
boundary of $D^{2} \times S^{1}$ with unit intersection with
$\lambda$, 
which we take to be $\partial D \times \{ 1\}$. 

We wish to perform surgery on $\Sigma \times S^{1}$ where $\Sigma$ is
a compact closed Riemann surface. Let $\Sigma_{0}=\Sigma/D_{1}^{2}
\cup \dots \cup D_{N}^{2}$ be the surface with 
the interiors of $N$ disjoint discs excised and, with obvious
notation, $\partial \Sigma_{0}= S^{1}_{1} \cup \dots \cup S_{N}^{1}$.
  We consider the manifold
$\Sigma_{0} \times S^{1}$. Denote the boundary curve
in $\Sigma_{0} \times S^{1}$ of the $i$'th excised disc in
$\Sigma_{0}$ by
$c_{i}= S_{i}^{1} \times \{1\} \subset S_{i}^{1} \times
S^{1}$. Likewise, denote $h_{i} = \{1\} 
\times S^{1} \subset S_{i}^{1} \times S^{1}$. 

Clearly we can
regain $\Sigma \times S^{1}$ by glueing solid tori to all of the boundaries
of $\Sigma_{0} \times S^{1}$ where we simply identify the $c_{i}$ with
the meridian $\mu_{i}$ and $h_{i}$ with the longitude $\lambda_{i}$ of the
$i$'th solid torus at the $i$'th boundary. 

More generally we could glue in the $N$ solid tori with the
identification, a homeomorphism $f$,
\be
f_{*}: \left( \begin{array}{c}
\mu_{i}\\
\lambda_{i} \end{array}\right) \longrightarrow \left(\begin{array}{rr}
a_{i} & b_{i}\\
-r_{i} & -s_{i} \end{array} \right)\, . \, 
\left( \begin{array}{c}
c_{i}\\
h_{i} \end{array}\right), \;\;   b_{i}r_{i}=1 + a_{i}s_{i}
\label{ab}
\ee
so that, in homology,
\bea
\mu_{i} & =&  a_{i}\, c_{i} + b_{i} \, h_{i} \nonumber \\
\lambda_{i} & = & -r_{i}\, c_{i} - s_{i}\, h_{i} \label{hol-1}
\eea
which reads, $\mu_{i}$ wraps $b_{i}$ times around
$h_{i}$ and $a_{i}$ times about $c_{i}$ while $\lambda_{i}$ wraps
$s_{i}$
times around $-h_{i}$ and 
$r_{i}$ times around $-c_{i}$. The image of $\{ 0 \} \times
S^{1}$ is called the singular fibre. Inverting the
relationship (\ref{hol-1}) we have
\bea
c_{i} & =&  -s_{i}\, \mu_{i} - b_{i} \, \lambda_{i} \nonumber \\
h_{i} & = & r_{i}\, \mu_{i}  + a_{i}\, \lambda_{i} \label{hol-2}
\eea
The manifolds that have just been
created, $M[g, \, (a_{1}, b_{1}), \dots , \,
(a_{N}, b_{N})]$, are Seifert manifolds but with non-normalised
Seifert invariants (so that $b_{i}$ is not necessarily smaller than
$a_{i}$).

The $S^{1}$ action (\ref{S1-action}) is designed to coincide with the
wrapping of
$h_{i}$ on $\partial V_{(a_{i}, b_{i})}$. To see this in detail let the
coordinate on $h_{i}$ be $\theta_{i}$ then by (\ref{hol-2}) the map
$S^{1} \rightarrow T^{2}$ with coordinates $(\phi_{i}, \, \psi_{i})$
on $T^{2}$ sends $\theta_{i}$ to $(r\theta_{i}, \, a_{i}\theta_{i})$
and the dual 1-form (\ref{dual-1-form}) pulls back to
$d\theta_{i}$.
Notice that this means that the solid tori $V_{(a_{i},b_{i})}$ come
complete with their surgery data, that is one glues the solid torus to
the rest of the manifold with the data (\ref{solid}) which is used in
(\ref{hol-1}) and (\ref{hol-2}).

As an example take $M=S^{2} \times S^{1} = \left( D^{2}
\cup D^{2}\right) \times S^{1}$ and take out the right hand $D^{2}
\times S^{1}$ (leaving us with another solid torus namely the left $D^{2}
\times S^{1}$) now glue back according to (\ref{ab}). The 3-manifold obtained
in this way is the Lens space $L(b,a)$ and in particular
$S^{3}=L(1,0)$ is obtained with $f$ given by
\be
\left(\begin{array}{cc}
0 & 1\\
-1 & 0\end{array} \right)
\ee

The prescription (\ref{ab}) is not the one required when one takes out
the tubular neighbourhood of a knot or link in $S^{3}$ and then glues
back\footnote{Rather, one uses instead a homeomorphism $\widehat{f}$,
$
\left(\begin{array}{cc}
b & -a\\
r & -s \end{array} \right)= \left(\begin{array}{cc}
a & b\\
s & r \end{array} \right) . \left(\begin{array}{cc}
0 & -1\\
1 & 0\end{array} \right) \nonumber
$ 
which first undoes the first glueing to get $S^{3}$ from $S^{2} \times
S^{1}$. In this case we have $
\widehat{f}_{*}(\mu) = b.c + a.h$.}.

\subsection{Fractional Monopole Bundles, Flat Connections and Surgery}

Now we would like to provide connections for the principal bundle
structure of the Seifert 3-Manifold as well as connections on bundles
over $M$. In the first case we wish to provide smooth
1-forms on the Seifert 3-Manifold obeying the 
usual conditions.

A natural connection one form on $V_{(a_{i},b_{i})}$ is (there is no
sum over a repeated index unless explicitely shown)
\bea
\sigma_{i} & = & if(\rho_{i})\, \left( b_{i}\, d\phi_{i} -
  s_{i} \, d\psi_{i}\right) + 
i\frac{1-f(\rho_{i})}{a_{i}}\,  d\psi_{i} \nonumber \\
d\sigma_{i} &=& i \frac{b_{i}}{a_{i}} df(\rho_{i}) \,
\wedge \, \left( a_{i} 
  d \phi_{i}  - r_{i} d
  \psi_{i}\right) \label{alpha}
\eea
where $f(0)=0$ and $f(1)=1$. The $\sigma_{i}$ satisfy
\be
\iota_{\xi_{i}}.\,\sigma_{i} = i, \;\;\; \mathrm{and} \;\;\;
\iota_{\xi_{i}}\, .\, d \sigma_{i} = 0.
\ee
The
first Chern class can be determined by integrating over the disc in
$V_{(a_{i},b_{i})}$ defined by $\psi =0$,
\be
c_{1} = \frac{1}{a_{i}} \frac{1}{2\pi i} \int_{D} i b_{i}\, df(\rho_{i})
\wedge d \phi_{i} 
= \frac{b_{i}}{a_{i}}
\ee
Note that
\be
\int_{V_{(a_{i},b_{i})}} \sigma_{i} \wedge d\sigma_{i} =  (2\pi i)^{2} \,
\frac{b_{i}}{a_{i}} 
\ee

If
one adds $i\, n_{i} f(\rho_{i}) (a_{i}  d\phi_{i} - r_{i} d\psi_{i})$
then we have a connection 
with $c_{1}= b_{i}/a_{i} + n$.

The holonomies for this connection along the meridian and longitude
(that is at $\rho =1$) are
\be
\mathrm{hol}_{\, \sigma_{i}}(\mu_{i}) = \exp{\left(2\pi i b_{i}\right)} =1, \;\;\;
\mathrm{hol}_{\, \sigma_{i}}(\lambda_{i}) = \exp{\left(-2\pi i s_{i}\right) } =1
\ee
so looking from the outside, as far as the boundary is concerned, one is
dealing with a flat connection. Consequently, if we glue the $i$-th
solid torus into $\Sigma_{0}$ with (\ref{hol-2}) 
and demand that the holonomy match we may extend the connection into
$\Sigma_{0}$ as a flat connection.  From our previous discussion the
extension into $\Sigma_{0} \times S^{1}$ is as
$d\theta$. Consequently, we define a 
continuous global 1-form 
$\kappa$ which on $\Sigma_{0} \times S^{1}$ is $d\theta$ and
\be
\left. \sigma\right|_{V(a_{i}, b_{i})} = \sigma_{i}
\ee
With any suitable choice of $f(\rho_{i})$, so that all its derivatives
vanish at $\rho_{i}=1$, $f^{(n)}(1)=0$ when $n \geq 1$, we obtain a
smooth connection one form. Indeed
with such a choice the curvature 2-form, $d\sigma_{i}$, vanishes at
$\rho_{i}=1$. With these choices $\sigma$ is a well defined smooth
connection 1-form, such that
\be
\int_{M} \sigma \wedge d\sigma = \sum_{i=0}^{N}
\int_{V_{(a_{i},b_{i})}} \sigma_{i} \wedge d\sigma_{i} = (2\pi i)^{2}
\left( b_{0} + \sum_{i=1}^{N} \frac{b_{i}}{a_{i}}\right) 
\ee
Notice that, in this way, we have defined a `global' principal bundle
structure on $M$. 

This bundle description can be made to be trivial
away from the fibres over the orbifold points, since we may choose $f$
to be one almost all the way into the center of the disc so that $d\sigma_{i}$
eventually has delta function support at the orbifold point. In
particular we have, suggestively in that limit, 
\be
\frac{d\sigma}{2\pi i} = b_{0}\, \delta(x_{0}) + \sum_{i=1}^{N}
\frac{b_{i}}{a_{i}}\, \delta(x_{i}),
\ee
with the $\delta(x_{i})$ being 2-form de-Rham currents.

\subsection{Holomorphic Description and Connections on
  $\mathcal{L}_{i}^{\otimes n_{i}}$}\label{Holomorphic}

Now we wish to connect the surgery prescription with that of complex
line V-bundles of the previous section.

Let $U(1)$ act on $\mathbb{C}$ by the character $e^{i\theta} \mapsto
e^{-in\theta}$. There are associated complex line V-bundles
$\mathcal{L}_{i}^{\otimes nb_{i}} $ over 
the orbifold $D^{2}/\mathbb{Z}_{a}$
\be
\mathcal{L}_{i}^{\otimes nb_{i}} = V_{(a_{i},b_{i})} \times_{n}
\mathbb{C}\label{assoc-bundle} 
\ee
meaning that $\mathcal{L}_{i}^{\otimes b_{i}} $ is the quotient of
$V_{(a_{i},b_{i})} \times
\mathbb{C}$ according to $(p, w).e^{i\theta} = (p.e^{i\theta},
e^{im\theta}.w) $. As before we can use the $S^{1}$ action to set
$\psi=0$ but we are left with a $\mathbb{Z}_{a}$ action
\be
(z, w) \simeq (\zeta^{r}.z, \zeta^{n}.w)
\ee
and $z=\rho \, . e^{i\phi}$. On using $\zeta^{b_{i}}$ as the generator rather
than $\zeta$ we get $(z, w) \simeq (\zeta .z, \zeta^{n b_{i}}.w)$ which
agrees with (\ref{local-L}).
The associated bundle construction (\ref{assoc-bundle}) allows us to
identify a holomorphic connection on the line bundle given the
connection (\ref{alpha}). We are tasked to make the identification
\be
(z,t, w) \simeq (z.\ex{ir_{i}\theta}, t.\ex{ia_{i}\theta}, \ex{-in\theta}.w)
\ee
which we do by taking $\ex{i\theta}=t^{-1/a_{i}}$ (and we still have to make the
identification under $\mathbb{Z}_{a_{i}}$). This provides us with a map
from $V_{(a_{i},b_{i})}$ to $\mathcal{L}_{i}^{\otimes n_{i}b_{i}} $ given by 
\be
(\rho_{i}\exp{i\phi_{i}}, \exp{i\psi_{i}})
\stackrel{\tau}{\longrightarrow} \left(
\rho_{i}.\exp{(i\phi_{i}-ir_{i}\psi_{i}/a_{i})},
\exp{(in_{i}\psi_{i}/a_{i})} \right) =(z_{i}, w_{i})
\ee

The connection (\ref{alphai}) on $\mathcal{L}_{i}^{\otimes n_{i}b_{i}} $
pulls back as
\be
\tau^{*}\left(\alpha_{i}^{\otimes n_{i}b_{i}}\right) & =& i n_{i} b_{i}
g(\rho_{i}^{2})(d\phi_{i}-r_{i} d\psi_{i}/a_{i}) + i
n_{i}d\psi_{i}/a_{i} + n_{i}b_{i} g(\rho_{i}^{2}) d \ln{\rho_{i}}
\ee
If we set $g(\rho_{i}^{2})= f(\rho_{i})$, which we do, then we have
the equality
\be
\tau^{*}\left(\alpha_{i}^{\otimes n_{i}b_{i}} \right)& = & n_{i} \sigma_{i} + d
\Lambda(\rho_{i})\label{form} 
\ee
We are really interested in the `classes' that these forms represent
and so we simply substitute $\alpha_{i}^{\otimes n_{i}b_{i}}$ with
$n_{i} \sigma_{i}$. In particular we have that the curvature 2-forms
agree,
\be
\tau^{*}\left(d\alpha_{i}^{\otimes n_{i}b_{i}}\right) = n_{i}
d\sigma_{i}
\ee

\section{Chern-Simons Theory on a Seifert 3-Manifold}

The Chern-Simons action is
\be
S = \frac{1}{4\pi}\int_{M}\Tr \left({\mathbb{A}d\mathbb{A} +
    \frac{2}{3}\mathbb{A}^{3}} \right)
\ee
where $\mathbb{A}$ is a connection on a trivializable (and
trivialized) $G$-bundle over $M$.

We consider the class of 3-manifolds $M$ as described in the previous
section, namely circle bundles $S(\mathcal{L})$ of holomorphic line V bundles
$\mathcal{L}$ over an orbifold $\Sigma_{V}$. As in \cite{BT-Seifert}, 
we take the gauge group $G$ to be a compact, semi-simple, connected
and simply connected Lie group.

Given the principal bundle structure $\kappa$ one can decompose fields
in a Fourier series along the fibre direction as done previously
\cite{BT-S1} and \cite{BT-Seifert}. One cannot completely follow the derivation
in those papers directly for reasons that we explained in the
Introduction though we will try to follow it as closely as possible.

Our conventions in \cite{BT-Seifert} had the vector field
generating the $U(1)$ action denoted by $\xi$ and the real dual 1-form
$\kappa$ satisfying
\be
\iota_{\xi}\kappa =1, \;\;\; \iota_{\xi} d\kappa = 0
\ee
with 
\be
d\kappa = - c_{1}\left(\mathcal{L}_{M} \right) \, \omega, \;\;\;
\int_{M} \kappa \wedge d\kappa = - c_{1}\left(\mathcal{L}_{M} \right), \;\;\;
\mathrm{where} \;\;\;
\int_{\Sigma_{V}} \omega =1 \label{convention}
\ee

We can achieve this be setting
\be
\kappa = \frac{1}{2\pi i } \sigma, \;\;\;
\left. \kappa\right|_{V(a_{i}, b_{i})}
= \kappa_{i} = \frac{1}{2\pi i } \sigma_{i}
\ee
and by understanding that the fibre has length 1 (rather than
$2\pi$). The minus sign in (\ref{convention}) implies that we are
using the opposite orientation for the Seifert 3-manifold $M$ to that
in previous sections. 

Now we decompose fields as
\be
\mathbb{A} = A + \kappa \phi
\ee
with $\iota_{\xi}A=0$ so that $A$ is a horizontal field with respect to
this fibration and $\phi$ is the component that lies along the
fibre. Note that both $A$ and $\phi$ are anti-Hermitian.

With this decomposition the Chern-Simons action becomes,
\be
S_{CS}[\mathbb{A}]= \frac{1}{4\pi}\int_M \Tr\left(
A \wedge \kappa \wedge L_{\phi}
\, A + 2\phi \, \kappa \wedge d\, A +
\phi^{2}\, \kappa\wedge d \, \kappa \right)\;\;.
\label{sm0}
\ee
The Lie derivative is denoted by
$L_{\xi}=\{ \iota_{\xi} ,\, d\}$
and for the covariant Lie derivative we set $L_{\phi}= L_{\xi}+ [ \phi,\, $

\subsection{Background Gauge Fields and Patches}

We know from the outset that once we try to impose the condition that
$\phi$ only takes values in the Cartan subalgebra that we will have to
sum over all possible non-trivial Abelian bundles that are `liberated'
in this procedure. In anticipation of this we will work
directly with a connection plus a background Abelian connection
\be
\mathbb{A} \rightarrow \mathbb{A} + \mathbb{A}_{B}
\ee
where $\mathbb{A}$ is a Lie algebra valued 1-form and $\mathbb{A}_{B}$ will be specified in the next section. To
explain where these background fields come from, recall that firstly
we set $\iota_{\xi}d \phi =0$ with a well defined gauge transformation
and then we follow this by diagonalising $\phi$ with a `time'
independent gauge transformation $\iota_{\xi}d g=0$ which is,
necessarily, singular. All of this is now happening on $\Sigma_{V} $
 and so the singular gauge transformations give rise to non-trivial
 bundles on $\Sigma_{V} $ so that 
one should be dealing with the connections we called $\alpha$ in
Section \ref{Sections}. However, we need to pull those back to our
3-manifold $M$ as in Section \ref{Holomorphic} and modulo a couple of
caveats that pull back (a sum of multiples of the $\kappa_{i}$) is our
background connection.

Given
that all the non-trivial bundles are encoded in the $\mathbb{A}_{B}$
we demand that $\mathbb{A}$ is a smooth globally defined form
(actually section). As the background is fixed gauge transformations
act as follows 
\be
\mathbb{A}^{g} = g^{-1}\mathbb{A}g + g^{-1}dg + g^{-1} \mathbb{A}_{B}g
- \mathbb{A}_{B}
\ee

Next we impose the gauge condition that $\phi$ is constant along the
fibre $\iota_{\xi}d\phi =0$. The variation of this condition involves
the operator
\be
L_{\phi + \phi_{B}}
\ee
where $\phi_{B}= \iota_{\xi}\mathbb{A}_{B}$ and it is this operator
that appears in the ghost determinant.

\subsection{Abelianization on a Seifert Manifold}

As we still have gauge invariance under those gauge transformations $g$
that satisfy $\iota_{\xi} dg=0$ we would like to Abelianize the field
$\phi$, that is set 
$\phi^{\lk}=0$ where we have decomposed the Lie algebra $\lg = \lt
\oplus \lk$ into a Cartan subalgebra and root spaces. If we do so then
we must follow this by summing over all
available line V bundles on the orbifold $\Sigma_{V}$. In previous
works on Abelianization in Chern-Simons theory this amounted to a sum
over one integer. The reason for that is that we had previously
considered Seifert 3-manifolds $M$ (the $\mathbb{Q}$HS$[g]$-manifolds
of \cite{BT-Seifert}) on which every line V bundle on the
base orbifold could be given as a tensor power of some unique line V
bundle $\mathcal{L}_M$. We are certainly far away from that
situation in the present context where we will need to sum over all
possible line V bundles.

To sum over all of these possibilities we add to the connection
$\mathbb{A}$ an Abelian background connection $\mathbb{A}_{B}$. The
Chern-Simons action goes over to,
\be
S_{CS}[\mathbb{A}+\mathbb{A}_{B} ]= S_{CS}[\mathbb{A}]
+\frac{1}{4\pi}\int_{M} \Tr{\left(\mathbb{A}_{B}d\mathbb{A}_{B} +
   2 \mathbb{A}\wedge F_{B}+ 
    2\mathbb{A}^{2} \mathbb{A}_{B} \right)} \label{cs-back}
\ee 
The last term only involves the charged components of the connection
$\mathbb{A}$ so that, in particular, it does not involve the gauge
fixed $\phi$. One may wonder why it is that $2 \mathbb{A}\wedge F_{B}$
appears in the action rather than $\mathbb{A}\wedge d\mathbb{A}_{B} +
d\mathbb{A} \wedge \mathbb{A}_{B}$  as, even though these two only
differ by an exact term $ 2 \mathbb{A}\wedge F_{B} = \mathbb{A}\wedge
d\mathbb{A}_{B} +
d\mathbb{A} \wedge \mathbb{A}_{B} - 
d(\mathbb{A} \wedge \mathbb{A}_{B})$, for singular forms a naive application of
Stokes theorem is not correct. Actually the Chern Simons Lagrangian is
not invariant under a gauge
transformation
\be
CS(\mathbb{A}^{g}) = CS(\mathbb{A}) + d \Tr{\left(\mathbb{A}^{g} \wedge
   g^{-1} dg\right)} - \frac{1}{3}   \Tr{\left(g^{-1}dg\right)^{3}}
\nonumber 
\ee
so, ignoring the winding number, we really should use
\be
CS(\mathbb{A}) = CS(\mathbb{A}^{g})- d \Tr{\left(\mathbb{A}^{g} \wedge
  g^{-1}  dg\right)} \nonumber
\ee
the $\mathbb{A}^{g}$ are our new `quantum' fields and $g^{-1}dg$ is essentially
$\mathbb{A}_{B}$.

The background bundles which are available to us are all of
those that can appear in (\ref{line-bundle-decomp}) 
\be
\mathcal{L}= \mathcal{L}_{0}^{\otimes n_{0}} \otimes
\mathcal{L}_{1}^{\otimes n_{1}} \otimes \dots \otimes 
\mathcal{L}_{N }^{\otimes n_{N}}\label{line-bundle-decomp-2}
\ee 
In the $i$-th patch the connection form $\sigma_{i}$ is that for $V_{(a_{i},b_{i})}$
(or equivalently $\mathcal{L}_{i}^{\otimes b_{i}}$). We would like to
represent connections of $\mathcal{L}_{i}^{\otimes n_{i}}$ but it seems that
the best that we can do is have connections for
$\mathcal{L}_{i}^{\otimes n_{i}b_{i}}$. In order to deal with this
situation we use (\ref{line-bundle-decomp-2}) as follows
\be
\mathcal{L}_{i}^{\otimes n_{i}r_{i}b_{i}} = \mathcal{L}_{i}^{\otimes
  n_{i}} \otimes \mathcal{L}_{0}^{\otimes s_{i} n_{i}}
\ee
from which we deduce that the connection that we require on
$\mathcal{L}_{i}^{\otimes   n_{i}}$ pulls back to $n_{i}r_{i}
\sigma_{i}- n_{i}s_{i} \sigma_{0}  $ or, somewhat more correctly, the
curvature 2-forms satisfy
\be
n_{i}[d\sigma_{i}] = n_{i}r_{i}[d\sigma_{i}] - s_{i}n_{i}[d\sigma_{0}]
\ee

The line bundles that appear live inside the gauge bundle that we are
considering, so that the background connection is taken to be
\bea
\mathbb{A}_{B} & = &  \mathbf{n}_{0} \sigma_{0} +
\sum_{i=1}^{N}\mathbf{n}_{i} \left( r_{i} 
\sigma_{i}- s_{i} \sigma_{0} \right) \nonumber \\
&=& 2\pi i \left( \mathbf{n}_{0} \kappa_{0} +
\sum_{i=1}^{N}\mathbf{n}_{i} \left( r_{i} 
\kappa_{i}- s_{i} \kappa_{0} \right)\right)
\eea
where the $\mathbf{n}$ are Hermitian. As the components of
$\mathbf{n}_{0}$ range over the integers, and as we will sum over
these, we can shift to absorb the $s_{i} \mathbf{n}_{i}$. With this
understood, and with an abuse of notation, we write the background as
\be
\kappa \, \phi_{B} = 2\pi i\left(\mathbf{n}_{0} \kappa_{0} +
\sum_{i=1}^{N}\mathbf{n}_{i} r_{i}\kappa_{i}
\right)
\ee


It is usually appropriate for an Abelian theory to write
\be
\frac{1}{4\pi}\int_{M} \Tr{\left(\mathbb{A}_{B} d\mathbb{A}_{B}\right)}
= \frac{1}{4\pi}\int_{X}
\Tr{\left(F_{B} \wedge F_{B}\right)}
\ee
where $X$ is a 4-manifold that bounds $M$. Recall that $M$ is itself
the unit circle V bundle, $S(\mathcal{L}_{M})$ of some line V bundle
$\mathcal{L}_{M}$. We are fortunate in that
there is a natural $X$ available to us, namely we take $X$ to be the
unit disc bundle $D(\mathcal{L}_{M})$ whose boundary is
$S(\mathcal{L}_{M}) \equiv M$. Even though though the disc bundle is
itself singular one could follow this through
\cite{OUYANG}, however, we 
are in the even happier situation that we are able to determine the
left hand side directly, which we now proceed to do. 

We can evaluate the second Chern-Simons contribution in
(\ref{cs-back}) as follows (with $r_{0}=1$)
\be
\frac{1}{4\pi}\int_{M} \Tr{\left(\mathbb{A}_{B} d\mathbb{A}_{B}\right)}
= - \pi \sum_{i=0}^{N} r_{i}^{2}\int_{V_{(a_{i},b_{i})}}
\Tr{\mathbf{n}_{i}^{2}}\, \kappa_{i} \wedge d\kappa_{i}  =  \pi
\sum_{i=0}^{N} 
\frac{r_{i}^{2}b_{i}}{a_{i}} 
\Tr{\mathbf{n}_{i}^{2} } \label{kdk}
\ee

For a simply connected group $G$,  $ \pi \Tr{\mathbf{n}_{i}^{2}}$ is an
element of $2\pi \mathbb{Z}$. The order of the normal point is 1
$(a_{0}=1,\, b_{0}=1)$ so
that the exponential of that term gives unity and so may be
neglected. Furthermore, on replacing $r_{i}b_{i}=1
+ a_{i}s_{i}$ in (\ref{kdk}), the only terms that will contribute in the
exponential are
\be
 \pi
\sum_{i=1}^{N} 
\frac{r_{i}}{a_{i}} 
\Tr{\mathbf{n}_{i}^{2} } \label{kdk2}
\ee 
For a non-simply connected group one will also have to take into
account signs that depend on the length of each of the $\mathbf{n}_{i}$.

In the second last term of (\ref{cs-back}) only the $\phi$ component of
$\mathbb{A}$ is present as $F_{B}$ is horizontal,
\be
\frac{1}{2\pi} \int_{\left(D_{i}\times S^{1}\right)/\mathbb{Z}_{a_{i}}}
\Tr{\left( \kappa \phi \wedge F_{B}\right)} =  -i
\frac{b_{i}r_{i}}{a_{i}}\Tr \phi \mathbf{n}_{i} 
\ee
and we have made use of the fact that when we integrate over Cartan
valued $A$ we have a delta function constraint on $\phi$ which implies
it is constant and 
\be
\left. \kappa\right|_{D_{i}} = d\theta + \beta_{i} 
\ee
and the integral on the fibre for $d\theta$ is one.

The last piece of the puzzle is the
\be
\frac{1}{2\pi}\int_{M} \Tr{\mathbb{A}^{2} \mathbb{A}_{B} } =
\frac{1}{4\pi}\int_{M} \Tr{A\wedge\kappa \wedge [\phi_{B}, A]  }
\ee
term where $\phi_{B}= \iota_{\xi}\mathbb{A}_{B}$. This piece appears
in the determinants that we have still to evaluate.

\subsection{Collecting Terms in the Action}

The total action becomes
\bea
ik S_{CS}& = &\frac{ik}{4\pi}\int_{M}   \Tr\left(
A \wedge \kappa \wedge L_{\left(\phi + \phi_{B}\right)}
\, A + 2\phi \, \kappa \wedge d\, A +
\phi^{2}\, \kappa\wedge d \, \kappa \right) \nonumber \\
& &  + i k \sum_{i=1}^{N} \frac{r_{i}}{a_{i}} \Tr{\left( -i b_{i}\phi 
    \mathbf{n}_{i} + \pi \mathbf{n}^{2}_{i}\right)} + k 
\Tr{\left( \phi \mathbf{n}_{0}\right)}
\label{sm1}
\eea
Clearly integrating over $A^{\lt}$ gives us the condition that
$d \left( \kappa \phi\right) =0$ which together with the gauge
condition on $\phi$ implies that $\phi$ is constant, $d\phi=0$. Now
that $\phi$ is constant and noting that
\be
\int_{M} \kappa \wedge d\kappa = - c_{1}(\mathcal{L}_{M})
\ee
we may write
\be
\frac{ik}{4\pi}\int_{M} \kappa \wedge d\kappa   \Tr{\left(
    \phi^{2}\right)} =- \frac{ik}{4\pi} c_{1}(\mathcal{L}_{M})\Tr{\left(
    \phi^{2}\right)} 
\ee

Consequently the partition function becomes
\be
Z_{CS} = \sum_{\mathbf{n}_{0} \in \mathbb{Z}}\left(\prod_{i=1}^{N}\sum_{\mathbf{n}_{i}=1}^{a_{i}-1}\right) \int_{\lt}d\phi\, 
\frac{\Det_{\Omega^{0}(M, \lk)}{\left(iL_{\phi + 
    \phi_{B}}\right)} }{\sqrt{\Det_{\Omega^{1}_{H}(M, \lk)}{\left(* \kappa
    \,\wedge \, 
    L_{\phi +
    \phi_{B}}\right)}}}\, . \,  \exp{\left( ikI(\phi, \mathbf{n})\right)}\label{part-kI}
\ee
where
\be
I(\phi, \mathbf{n})=  \sum_{i=1}^{N}
\frac{r_{i}}{a_{i}} \Tr{\left( -i b_{i}\phi  
    \mathbf{n}_{i} + \pi r_{i}\mathbf{n}^{2}_{i}\right)} +
\Tr{\left( \phi \mathbf{n}_{0}\right)} - \frac{1}{4\pi}
c_{1}(\mathcal{L}_{M})\Tr{\left( \phi^{2}\right)} \label{kI}
\ee

\section{One Loop Effects and the Kawasaki Index Theorem}

We borrow heavily from the calculations in \cite{BT-Seifert}. In order
to make contact with that work we will need to explain, along the way,
how working locally
mimics the global calculations there. Furthermore, we need to take into
account that $\phi_{B}$ unlike $\phi$ is not constant on
$\Sigma_{V}$. Lastly, one needs to note that the Kawasaki index theorem
tells us that the number of holomorphic sections of the line V bundle
only depends on the desingularisation $|\mathcal{L}|$ over the smooth 
manifold $\Sigma \equiv |\Sigma_{V}|$ of the 
holomorphic line V bundle $\mathcal{L}$ over $\Sigma_{V}$,
\be
\chi(\Sigma_{V}, \mathcal{L}) = 1-g +  \deg{\left( \mathcal{L} \right)}
\ee
where $\deg{\left( \mathcal{L} \right)}= c_{1}{\left( |\mathcal{L} |\right)}$.

Firstly we write the ratio of determinants in terms of Fourier modes
as sections of powers, $\mathcal{L}_{M}^{\otimes n}$ of the line V bundle that
defines $M$,
\be
\frac{\Det_{\Omega^{0}(M, \lk)}{\left(L_{\phi + 
    \phi_{B}}\right)} }{\sqrt{\Det_{\Omega^{1}_{H}(M, \lk)}{\left(* \kappa
    \,\wedge \, 
    L_{\phi +
    \phi_{B}}\right)}}}
\ee

We regularise both the absolute value and the phase of the ratio of
determinants as follows
\be
\sqrt{\Det{Q}} = \sqrt{\left|\Det{Q}\right|}\, \exp{\left(\frac{+i\pi}{2} \,
  \eta(Q) \right)}
\ee
where $
\eta(Q) = \frac{1}{2} \sum_{\lambda \in \mathrm{spec}(Q)} \mathrm{sign}(\lambda)$
\bea
\left|\Det{Q}\right|(s) &=& \exp{\sum_{\lambda\in \mathrm{spec}(Q)}
  e^{ s \Delta} \ln 
  |\lambda|} \label{absdet}
\\
\eta(Q, \, s) &=& \frac{1}{2} \sum_{\lambda\in \mathrm{spec}(Q)}
\frac{\mathrm{sign}(\lambda)}{|\lambda|^{s} }\exp{\left( s \Delta
  \right) } \label{etas}
\eea
for $\Delta$ the Laplacian of the twisted Dolbeault operator.

\subsection{The Absolute Value of the Determinants and Ray-Singer Torsion}\label{calc}

Had $\phi_{B}$ been constant then the regularisation would have led us
to consider \cite{BW}
\be
\chi\left(\Sigma, \mathcal{L}^{\otimes n}_{M}\right) + \chi\left(\Sigma,
  \mathcal{L}^{\otimes -n}_{M}\right) = 2-2g -N +
\sum_{i=1}\phi_{a_{i}}\left(n \right)
\ee
What $\phi_{a_{i}}(n)$ measures are the number of `honest' line bundles
in the tensor power $\mathcal{L}_{i}^{\otimes nb_{i}(\mathcal{L}_{M})}$, that
survive in the sum $\deg{\mathcal{L}_{M}^{\otimes n}}
+\deg{\mathcal{L}_{M}^{\otimes -n}}  $, and
these line bundles are, by construction, at the $i$'th orbifold
point. Note that we always have $\gcd{\left(a_{i}, \,
    b_{i}(\mathcal{L}_{M}) \right)} =1$ so that an honest bundle only
arises when $a_{i}|n$.

However, $\phi_{B}$ is not constant. As explained between (6.16) and
(6.19) in \cite{BT-CS} when dealing 
with a non constant $\phi$ the ratio of determinants takes the form of
an integral of the density representing the characteristic classes of
the Dolbeault operator and the, log of, the operator itself. Applying
that in the orbifold case leads us to objects of the form
\be
\int_{\Sigma_{V}}\mathcal{I}\left(R,\, F_{0}, \dots, F_{N} \right) \ln{M(\phi +
  \phi_{B})} \label{index-local}
\ee
in the effective action. Here $\mathcal{I}\left(R,\, F_{0}, \dots, F_{N} \right)$ is the local
density function of the characteristic classes and $M(\phi +
  \phi_{B})$ is essentially $\left. \Det_{\lk}\left(L_{\phi + \phi_{B}
      }\right)\right|_{S^{1}}$ (which varies over $\Sigma_{V}$).

We use the local decomposition
(\ref{line-bundle-decomp}) for line V bundles whose support is about
the specified points $x_{0}, \, x_{1}, \dots , x_{N}$ and we recall that
contributions to the index theorem are local to express
(\ref{index-local}) as
\be
\int_{\Sigma_{0}}\mathcal{I}\left(R,\, F_{0} \right) \ln{M(\phi +
  \phi_{B})} + \sum_{i=1}^{N} \int_{D_{i}/\mathbb{Z}_{a_{i}}}
\mathcal{I}\left(R,\, F_{i} \right) \ln{M(\phi + 
  \phi_{B})}
\ee
where $\Sigma_{0}$ is $\Sigma_{V}$ with the $N$ 
discs about the orbifold points removed. As the Kawasaki index comes
from the holomorphic Lefshetz fixed point formula there are
contributions coming from the orbifold points which we have
implicitly incorporated in the integrals over the
$D_{i}/\mathbb{Z}_{a_{i}}$. Indeed as we saw previously by appropriate
choice of $f$ in (\ref{alpha}), we can have delta function support for
the curvature 2-forms and thus `localise' the contribution to the
fixed points.

On each region $\phi_{B}$ is constant
\be
\left. \phi_{B} \right|_{U_{i}} = 2\pi i r_{i} \mathbf{n}_{i}
\ee
Over $\Sigma_{0}$ there are only smooth
line bundles which cancel out in the sum of degrees which leaves us
with the Euler characteristic which is $2-2g-N$ and following the
discussion in section 5.1 of \cite{BT-Seifert} this leads us to a
factor of
\be
T_{S^{1}}\left(\phi +2\pi i \mathbf{n}_{0}
\right)^{1-g-N/2} = T_{S^{1}}\left(\phi\right)^{1-g-N/2}
\ee
where $T_{S^{1}}(\varphi)$ is the Ray-Singer torsion on $S^{1}$ of a
constant connection $\varphi \, d\theta$ (and all connections are
gauge equivalent to such a connection). So, on $\Sigma_{0}$,
$M=T_{S^{1}}$ and
\be
T_{S^{1}}(\varphi) = \det _{\lk}{\left(1 - \Ad{\, \ex{\varphi}} \right)}
\ee
where the right hand side is a determinant on the $\lk$ part of the
Lie algebra $\lg$.

As we saw before there can also be contributions of honest line
bundles at the orbifold points (though we will have to consider the
orbifold points
to be `smoothed out'). Recall that the $\phi_{a_{i}}$ count those
line bundles over $D_{i}$ which cancel in the sum of Euler characteristics. Indeed
$\phi_{a_{i}}(n)$ arises as
\be
\phi_{a_{i}}(n)= 1 -\frac{1}{a_{i}}\left(
  b_{i}(\mathcal{L}_{i}^{\otimes nb_{i}}) + 
  b_{i}(\mathcal{L}_{i}^{\otimes - nb_{i}}) \right)
\ee
where the $1$ is the Euler characteristic of the disc and the second
term is $0$ if $a_{i}|n$ and one otherwise. When $a_{i}|n$
then line V bundles are line bundles and the second term vanishes
(honest line bundles drop out in the sum).

In any case, once more
following the discussion in section 5.1 of \cite{BT-Seifert} gives us
the factor
\be
T_{S^{1}}\left((\phi  +
  2\pi i r_{i}\mathbf{n}_{i})/a_{i} \right)^{1/2}
\ee

All together then we have that the absolute value is
\be
T_{M}(\phi; \, \mathbf{n}_{i}) =
T_{S^{1}}\left(\phi\right)^{1-g-N/2}. \prod_{i=1}^{N}T_{S^{1}}\left((\phi 
  + 2\pi i
  r_{i}\mathbf{n}_{i})/a_{i} \right)^{1/2} 
\ee

\subsection{The Phase of the Ratio of  Determinants and $\eta$ Invariants}\label{Phase}

We recall the regularised formulae for the phase with $\phi_{B}$ constant
and then take into account the fact that it is not so.

In section 5 of \cite{BT-Seifert} the phase is split into two pieces
one depending on the charges of the fields but not on the line  V
bundles defining $M$ while the second has dependence on
$\mathcal{L}_{M}$ but not on the smooth line bundles $V_{\alpha}$,
\be
\eta(L_{\phi+\phi_{B}},s) = \sigma(L_{\phi+\phi_{B}},V_{\lk}, s) +
\gamma(L_{\phi+ \phi_{B}},\mathcal{L}_{M}, s) \nonumber
\ee
where
\bea
& & \sigma(L_{\phi + \phi_{B}},V_{\lk}, s) \nonumber \\
& & \;\;\; =  - 2 \sum_{\alpha >0}\deg{(V_{\alpha})}|i
\alpha(\phi+ \phi_{B})|^{-s}-2 \sum_{\alpha >0}\deg{(V_{\alpha})}\sum_{n\geq
  1}(2\pi n+i 
    \alpha(\phi+\phi_{B}))^{-s} \nonumber\\
& & \;\;\;\;\;\;\;\;\;\; + 2 \sum_{\alpha >0}\deg{(V_{\alpha})}\sum_{n\geq
  1} (2\pi n- i \alpha(\phi+ \phi_{B}))^{-s}\label{etaV}
\eea
and
\bea
\gamma(L_{\phi+\phi_{B}},\mathcal{L}_{M},s) &= & - \sum_{n\geq 1}\sum_{\alpha >0} \;\;\;
    [\deg{(\mathcal{L}^{\otimes n}_{M})}-\deg{(\mathcal{L}^{\otimes 
        -n}_{M})} ]\,\nonumber\\
& & \;\;\;\; .\;\; \left[ (2\pi n+i
    \alpha(\phi+ \phi_{B}))^{-s} + (2\pi n-i
    \alpha(\phi +\phi_{B}))^{-s}\right]
\label{etaphi}
\eea
However, here it is not the case that the bundle dependence neatly
seperates. Both terms depend on the 
background gauge field and we are in danger of overcounting. It is
straightforward to see that one generates equivalent terms in
$\sigma(L_{\phi + \phi_{B}},V_{\lk}, s) $ and
$\gamma(L_{\phi+\phi_{B}},\mathcal{L}_{M},s)$ if one allows both to
have the background field dependence. As we have extracted the
background gauge fields we understand the 
field strength associated with the $V_{\alpha}$ in the above formula
to be $dA$, and that the background field dependence should be turned
off. With this understood the contribution to
(\ref{etaV}) is
\bea
& & \sigma(L_{\phi},V_{\lk}, s) \nonumber \\
& & \;\;\; =  - 2 \sum_{\alpha >0}\deg{(V_{\alpha})}|i
\alpha(\phi)|^{-s}-2 \sum_{\alpha
  >0}\deg{(V_{\alpha})}\sum_{n\geq 
  1}(2\pi n+i 
    \alpha(\phi))^{-s} \nonumber\\
& & \;\;\;\;\;\;\;\;\;\; + 2 \sum_{\alpha >0}\deg{(V_{\alpha})}\sum_{n\geq
  1} (2\pi n- i \alpha(\phi))^{-s}\label{etaV2}
\eea
In the limit as $s\rightarrow 0$
\be
\sigma(L_{\phi + \phi_{B}},V_{\lk}, s) =-2 \sum_{\alpha > 0}
\deg{\left(V_{\alpha} \right)}\left( 1 +
\frac{1}{\pi} i \alpha(\phi) \right) + \mathcal{O}(s)
\ee
The term
$\sum_{\alpha>0} \deg{\left(V_{\alpha} \right)}$
does not contribute to the phase, given our assumption that the group
is simply-connected, so we are left with 
\be
\sigma(L_{\phi + \phi_{B}},V_{\lk}, 0) \rightarrow - \frac{2i}{\pi} \sum_{\alpha > 0}
c_{1}\left(V_{\alpha} \right)
\alpha(\phi) 
\ee

In order to define and then evaluate (\ref{etaphi}) when $\phi_{B}$ is not
constant we must define what we mean by the right hand side of
\be
\deg{(\mathcal{L}^{\otimes n}_{M})}-\deg{(\mathcal{L}^{\otimes 
        -n}_{M})} =
2n.c_{1}(\mathcal{L}_{M}) - 2 \sum_{i=1}^{N}
\ldb\frac{n b_{i}(\mathcal{L}_{M})}{a_{i}}\rdb 
\ee
The first Chern character $c_{1}(\mathcal{L}_{M})=
c_{1}(\mathcal{L}^{\otimes b_{0}}_{0})
\oplus_{i=1}^{N} c_{1}(\mathcal{L}_{i}^{\otimes b_{i}})$ and each
summand is supported on the corresponding open set. The terms
involving the double bracket symbol 
$\ldb nb_{i}/a_{i}\rdb$ come from densities that have support on $U_{i}$.

In section 5, equation (5.26) of \cite{BT-Seifert} the phase
proportional to $c_{1}(\mathcal{L}_{M})$ is determined to be
\be
c_{1}(\mathcal{L}_{M})\sum_{\alpha >0} \left(
  \frac{1}{3} +\frac{1}{2\pi^{2}}\alpha(\phi)^{2}  \right) +
\mathcal{O}(s)
\ee
which now goes over to
\bea
& & \sum_{i=0}^{N}c_{1}(\mathcal{L}_{i}^{\otimes b_{i}})\sum_{\alpha >0} \left(
  \frac{1}{3} +\frac{1}{2\pi^{2}}\alpha(\phi + 2\pi i r_{i}\mathbf{n}_{i}
  )^{2}  \right) + 
\mathcal{O}(s) \nonumber \\
& &\longrightarrow  
\sum_{\alpha>0} \left(
  \frac{c_{1}(\mathcal{L}_{M})}{2\pi^{2}}\alpha(\phi)^{2}
  +\frac{2i}{\pi}\alpha(\phi)\alpha(\mathbf{n}_{0}) +
  \frac{2}{\pi}\sum_{i=1}^{N} \frac{b_{i}r_{i}}{a_{i}}\left[
i\alpha(\phi)\alpha(\mathbf{n}_{i}) - \pi r_{i}\alpha( \mathbf{n}_{i})^{2}
 \right]\right)\nonumber\\
& & \;\;\;\;\;\;\; + \dim{(G/T)}\, \frac{c_{1}(\mathcal{L}_{M})}{6} 
\eea

The determination of the phase coming from the double bracket symbol
is presented between (5.26) and (5.27) in \cite{BT-Seifert}. As one
can see there that calculation is done for each line V bundle
$\mathcal{L}_{i}$ independently and does not depend on $\phi$
consequently (5.27) there immediately goes over to
\be
2\sum_{\alpha >0}\sum_{n\geq 0} \sum_{\pm}\ldb\frac{nb_{i}}{a_{i}}\rdb
.\frac{1}{(2\pi n \pm i 
        \alpha(\phi + \phi_{B}))^{s}}= -2\dim{(G/T)}\,  s(b_{i}, a_{i})+
      \mathcal{O}(s)\nonumber 
\ee
without change.

Collecting all the contributions including one from the $T$ valued
fields we have (mod $ 4\mathbb{Z}$)
\bea
\eta(0) & = &   \sum_{\alpha >0} \left(
  \frac{c_{1}(\mathcal{L}_{M})}{2\pi^{2}}\alpha(\phi)^{2} -
  \frac{2c_{1}(V_{\alpha})}{\pi} 
  i\alpha(\phi) +\frac{2i}{\pi}\alpha(\phi)\alpha(\mathbf{n}_{0}) \right)\nonumber \\
&& \;\;\; +\sum_{\alpha >0}
  \frac{2}{\pi}\sum_{i=1}^{N} \frac{b_{i}r_{i}}{a_{i}}\left(
i\alpha(\phi)\alpha(\mathbf{n}_{i}) - \pi r_{i}\alpha( \mathbf{n}_{i})^{2}
 \right)+ \dim{G}\left(   \frac{c_{1}(\mathcal{L}_{M}) }{6} - 2
  \sum_{i=1}^{N}s(b_{i},  a_{i}) \right) \nonumber
\eea
so that we have finally determined the phase to be
\bea
-\frac{i\pi}{2} \eta(0) &=& 4\pi i \Phi(\mathcal{L}_{M}) -
\frac{ic_{\lg}}{4\pi} c_{1}(\mathcal{L}_{M}) 
  \Tr{(\phi^{2})} + \frac{ic_{\lg}}{2\pi}
\int_{\Sigma} \Tr \phi F_{A} \nonumber \\
 & & \;\;\;\; + c_{\lg} \Tr{\left(\phi \mathbf{n}_{0} \right)} + i c_{\lg}\sum_{i=1}^{N} \frac{b_{i}r_{i}}{a_{i}} \Tr{\left(
-i\phi \mathbf{n}_{i} + \pi r_{i}\mathbf{n}_{i} ^{2}
 \right)}\label{total-phase}
\eea

The term $\int_{\Sigma} \Tr \phi F_{A}$ should not really be
considered as we have already taken $\phi$ constant and all the
non-trivial bundle structure resides in the background fields. This is
unlike previous works on abelianisation where $\int_{\Sigma} F_{A}
\neq 0$. However, thinking of the gauge field $A$ in (\ref{sm1}) as a
background field, for the purposes of this calculation, then we indeed
get the appropriate shift in $k$ for this term too.

\subsection{The Partion Function}

The net effect of the phase is to give us the famous shift $k
\rightarrow k_{\lg}= k+ c_{\lg}$ as well as the framing term
\be
\Phi(\mathcal{L}_{M}) = - \frac{\dim{G}}{48}\left(
  c_{1}(\mathcal{L}_{M})  -  12
  \sum_{i=1}^{N}s(b_{i},  a_{i}) \right)
\ee

Consequently the partition function becomes
\be
Z_{CS} =  \sum_{\mathbf{n}_{0} \in
  \mathbb{Z}}\left(\prod_{i=1}^{N}\sum_{\mathbf{n}_{i}=1}^{a_{i}-1}\right)
\int_{\lt}d\phi\;\;\sqrt{T_{M}(\phi; \, \mathbf{n}_{i})} 
 . \; \exp{\left( 4\pi i\Phi(\mathcal{L}_{M})
    +ik_{\lg}I(\phi, 
    \mathbf{n})\right)}\label{part-kI2} 
\ee

There is still a large symmetry available to us. The first Chern class
is a rational number so we set\footnote{We are not claiming that there
exists an $\mathcal{L}_{0}$ for which $\mathcal{L}_{M}$ is the $d$'th
tensor power.} $c_{1}(\mathcal{L}_{M})=d/P$ where
$P=a_{1} \dots a_{N}$
\be
\phi \rightarrow \phi - 2\pi P \mathbf{s}, \;\;\; \mathbf{n}_{0}
\rightarrow \mathbf{n}_{0} + d \mathbf{s}
\ee
Just as in \cite{BT-Seifert} (2.4) one may use this symmetry to write the
partition function in various forms.
To write the partition function completely as a sum one
only needs to note that using the symmetry we may constrain the $\phi$
integrals to lie between zero and $2\pi P$ while performing the sum
over $\mathbf{n}_{0}$ sets 
\be
\phi = 2\pi \mathbf{n}/k_{\lg} \label{phi-is-int}
\ee
Note that if $c_{1}(\mathcal{L}_{M})=0$, there is still the symmetry
one must set $d=0$. 

On setting $\phi$ to be as in (\ref{phi-is-int}) then every occurrance
of the product $b_{i}r_{i}$ in the exponential in (\ref{part-kI2}) can
be taken to be unity thanks to an argument we have used a number of
times. With this substitution understood then these formulae agree well (up to an overall factor, which can be
determined) with \cite{Rozansky, Hansen, Hansen-Takata}.
\section{Odds and Ends}

It might seem that the introduction of the background fields changes
the fibre Wilson loop observables that we are able to
evaluate. However, this is not the case. Depending over which open set
$U_{i}$ we are the observable becomes, on abelianisation and noting
that $\phi$ is gauge fixed to be constant on the fibre,
\be
\Tr_{R}{\left(P \exp{\left(\oint \kappa (\phi + \phi_{B} )\right)}\right)} =
\Tr_{R}{\left( \exp{\left(
    \phi + 2\pi b_{i}^{*}\mathbf{n}_{i}\right)}\right)} =
\Tr_{R}{\left( \exp{\left(     \phi\right)}\right)} \label{obs}
\ee
This shows us that for such loops it is not important which smooth point
on the base they go through in the fibration $S^{1} \rightarrow M
\rightarrow \Sigma_{V}$. To evaluate the expectation value of products
of such knots one may simply insert the appropriate operators with
representations in (\ref{part-kI2}).

As explained in \cite{Beasley} and developed in detail for complex
Chern Simons theory in \cite{BT-complex-CS} one can use different
Seifert representations of the same manifold to obtain the invariants
of different knots.

The same arguments that we have given apply to other theories such as
$BF$ theory and Chern Simons theory with a complex gauge group.

{\bf Acknowledgements}

 The work of Matthias
 Blau is partially supported through the NCCR SwissMAP (The
 Mathematics of Physics) of the Swiss National 
Science Foundation. Keita Kaniba Mady would like to thank the Abdus Salam ICTP for the
 hospitality extended to him during his doctorate studies and the Ministry of
 Education of Mali for their financial support.  

\rnc{\Large}{\normalsize}

\end{document}